\documentclass[preprint,pre,showpacs,preprintnumbers,floatfix,amsmath,amssymb]{revtex4}

\usepackage{graphicx}
\usepackage{epsfig}

\begin{document}

\title{Transient polarization dynamics in a CO$_2$ laser}

\author{I. Leyva$^{1,2}$}
\email{ileyva@ino.it}
\author{E. Allaria$^{1}$}
\author{R. Meucci$^{1}$}

\affiliation{$^1$ Istituto Nazionale di Ottica Applicata, Largo E. Fermi 6, 50125 Florence, Italy}
\affiliation{$^2$ Universidad Rey Juan Carlos. c/Tulipan  s/n. 28933 Mostoles Madrid, Spain}

\begin{abstract}
We study experimentally and theoretically the polarization alternation during the switch-on
transient of a quasi-isotropic CO$_2$ laser emitting on the fundamental mode. The observed 
transient dynamics is well reproduced by means of a model which provides a quantitative discrimination between 
the intrinsic asymmetry due to the kinetic coupling of molecules with different angular momenta, and
the extrinsic anisotropies, due to a tilted intracavity window. Furthermore, the experiment provides a
numerical assignment
for the decay rate of the coherence term for a CO$_2$ laser.
\end{abstract}
 
 \maketitle

\section{Introduction}
\label{s:intro}

Laser dynamics is commonly studied considering the electric field
as a scalar variable, since in most systems the
polarization state is imposed by anisotropies of the cavity. For instance,
Brewster windows or gratings, generally used in gas lasers to
close the laser tube or to select a vibro-rotational transition,
impose a linearly polarized state of the laser emission.
However, in perfectly cylindrical laser cavities without any elements to
select a preferred polarization, the study of the dynamics includes the
necessity of considering the vector nature of the electric field.

Several theoretical works have been devoted to the study of the polarization dynamics
of the quasi-isotropic laser, showing the important role played by the material variables.
In particular, the degeneracy of the angular momentum states of the laser transition sublevels
has been considered as the coupling source between different polarization states.
Initial studies considered only stationary solutions \cite{Lamb67,Bakaev87}, but
more recently dynamical models have been developed
to explore the role of anisotropy due to the laser medium,
that from now on we will call intrinsic anisotropy
\cite{Puccioni87,Abraham95,Abraham96,Stephan98}. These
models predict a rich dynamics even for the simplest transition, $J=1 \rightarrow J=0$. Later versions 
of these models include also 
linear and circular cavity anisotropies that we call 
extrinsic \cite{Matlin95,Redondo97}, or different level structure \cite{Vilaseca}.

However, up to now few experiments have been performed on this subject.
Experiments carried out on noble
gas lasers reveal that in some systems it is necessary to
consider the dynamics of the matter variables to fully understand
the polarization features of an isotropic system \cite{Bakaev87,Puccioni89}. In other
cases, the observations could be explained by a nonlinear
coupling of the modes and residual cavity anisotropies
\cite{Taggiasco97,Labate98}.

In this work we present experimental and theoretical results for the polarization
competition dynamics in the transient state of a quasi-isotropic low-pressure CO$_2$ laser.
We show that the polarization dynamics during the switch-on is well
described by means of a model including optical coherences (intrinsic anisotropy)
and extrinsic linear anisotropies.
These results extend our previous work in which the coherences were simplified as
a parametric cross coupling between the matter polarization and electric fields \cite{Leyva01}, and provide
a quantitative comparison between the amount of anisotropy due to the coupling of molecular angular
momenta and that induced by an intracavity window.
Furthermore, the decay rate of the coherences,
defined  in Ref. \cite{Puccioni87}, is shown explicitly to be closer to the decay rate of the
population inversion rather than to the decay rate of the induced polarization.

\section{Experimental results}
\label{s:setup}

\begin{figure}
\centerline{\scalebox{0.4}{\includegraphics{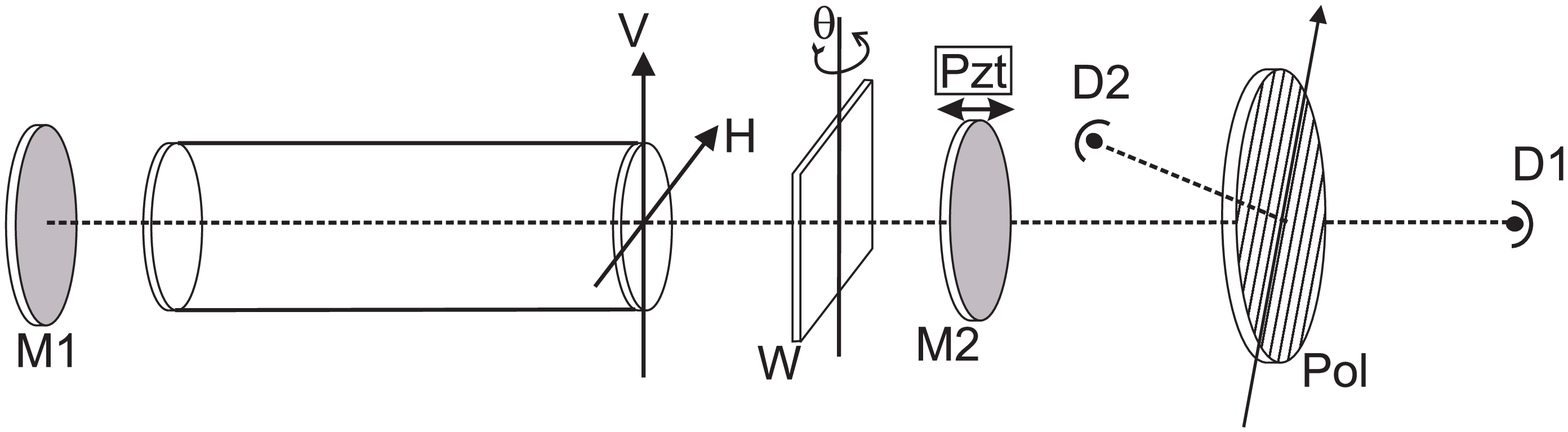}}}
\caption{Experimental setup:
M$_1$: total reflecting flat mirror,
W: additional intracavity window,
M$_2$: outcoupler mirror,
V: Vertical axis of the cavity,
H: Horizontal axis of the cavity,
Pzt: piezo electro translator,
Pol: wire grid polarizer,
D$_1$: fast detector for the vertical component,
D$_2$: fast detector for the horizontal component
}
\label{setup}
\end{figure}

The experiment has been performed using an unpolarized Fabry-Perot
cavity as shown in Fig. \ref{setup}.
A total reflective flat mirror (M$_1$, reflectivity R$_1$=1) and an outcoupler mirror
(M$_2$) with a reflectivity $R_2=0.914$ set the cavity length at $L=$75 cm.
A piezo translator (Pzt) is used to select the P(20) laser emission line and to
adjust the laser detuning.

The active medium, a mixture of $He$ (82$\%
$), N$_2$ (13.5$\%
$) and CO$_2$ (4.5$\%
$)
at a pressure of
25 mbar, is pumped by a DC discharge fixed at 6.1 mA when the threshold current
is 3 mA.

As we are interested in the transient dynamics, an intracavity chopper is used to
induce a switch-on event at a repetition rate of 200 Hz.

In order to control the linear anisotropies of the cavity, a ZnSe anti reflection
coated window (W) has been introduced, which can be tilted accurately
as shown in Fig.\ref{setup}.

The polarization state of the laser emission has been analyzed
by means of a wire grid polarizer (Pol), which has the property of
reflecting one linear polarization of the incident radiation and transmitting the
orthogonal one, with an extinction ratio of
1:180. The reflected and transmitted parts of the beam are directed to two
HgCdTe fast detectors (D$_1$,  D$_2$, 100 MHz bandwidth), whose sensitive
areas ($10^4$ $\mu $m$^2$) are much smaller than the beam size. Both signals
are recorded on a digital oscilloscope (Lecroy
LT423L) with $500$ MHz bandwidth.


In the stationary regime, we observe that the laser has two possible linear polarization directions
which are orthogonal as far as we can measure. In Fig. \ref{setup} these cavity eigendirections
are called H and V respectively.

\begin{figure}
\centerline{\scalebox{1.0}{\includegraphics{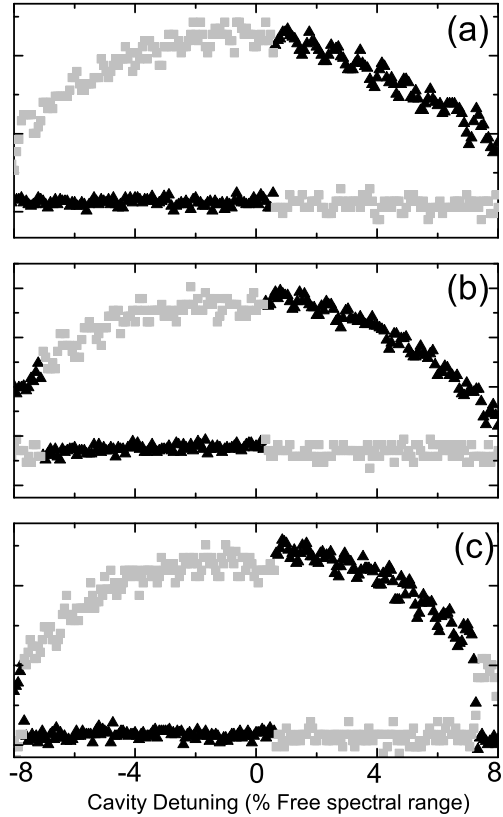}}}
\caption{Experimental intensity of both polarization components along a free spectral range
(Black triangles vertical polarization direction, gray squares horizontal polarization direction)
(a) quasi isotropic condition $\theta=0$
(b) $\theta =6^o$
(c) $\theta =9^o$.
It is to be noted that a Pzt adjustement is required  to recover the resonant condition for each $\theta$.}
\label{detuning_scan}
\end{figure}

\begin{figure}
\centerline{\scalebox{1.0}{\includegraphics{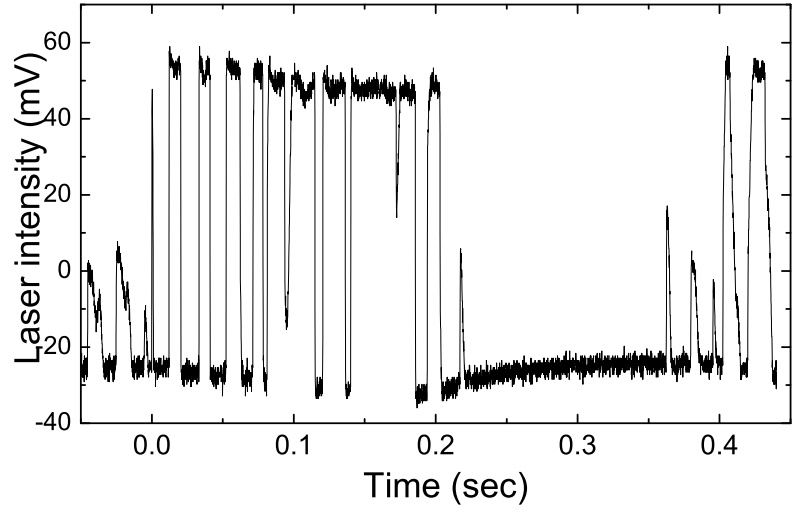}}}
\caption{ Intensity of vertical polarization component
when the cavity detuning is chosen at the transition point near the resonance.}
\label{flipflop}
\end{figure}

 The window W can rotate by an angle $\theta$ around the axis V. When $\theta=0$, the polarization direction is determined
by the detuning of the cavity, as usual \cite{Bakaev87,Matlin95,Eijkele95}. This is illustrated in
Fig.\ref{detuning_scan}(a), where we plot the laser intensity
along both eigendirections when the detuning is varied around the line center. A residual hysteresis 
around the transition is masked by the limited resolution of Pzt.
At the transition point near the line center  there is not a preferred polarization direction and
the laser flips spontaneously from one to other as shown in Fig. \ref{flipflop}.

Our goal is to study the polarization behavior in this bistable region.
In particular, we analyze of the switch-on transient state,
in which the total intensity presents relaxation oscillations.
In our system the pump strength imposes a relaxation frequency of 55 KHz.

\begin{figure}
\centerline{\scalebox{1.35}{\includegraphics{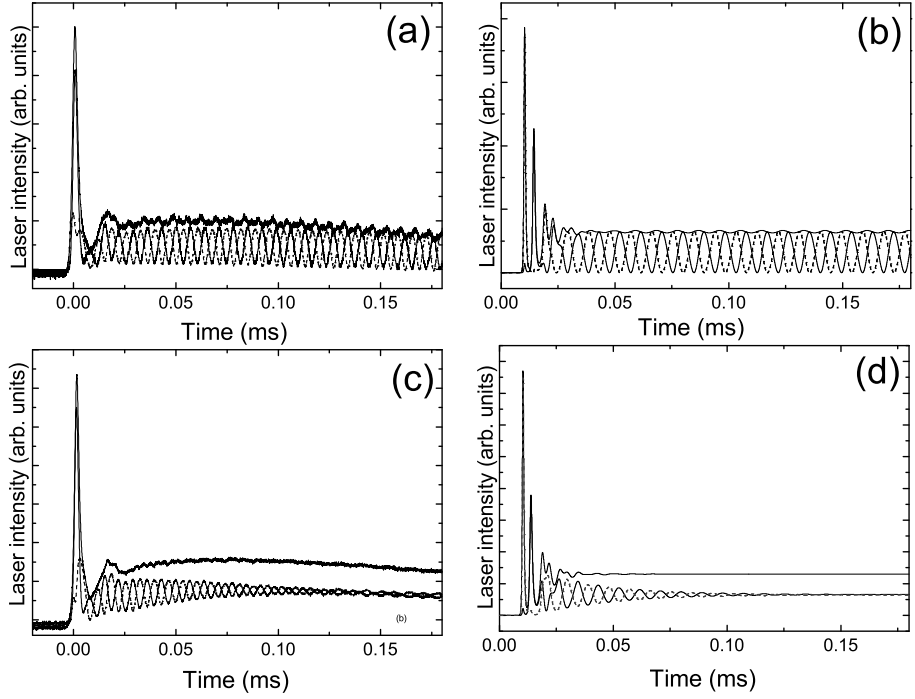}}}
\caption{Experimental time intensity profiles for $\theta=0$ and slightly different
detuning condition : (a) resonance, (c) out of resonance.
Numerical generated intensity profiles for $\theta$=0, $\alpha=\beta=0.01$ and : (b) $\delta=0$ ,
(d) $\delta=0.05$.}
\label{monomode}
\end{figure}

When the two polarized components are separated, they show oscillations
in relative antiphase, which do not appear on the total intensity (Fig.\ref{monomode}).
The amplitude of these oscillations depends on the angle between the
polarizer axes with respect to H-V, reaching a maximum when the analysis is performed at 45$^o$.

In Fig.\ref{monomode}(a) we report an example of these polarization oscillations,
when the cavity is tuned at the center of the line.
The total laser output intensity (thick solid line) is displayed together with its two orthogonally polarized components
analyzed at 45$^o$ (thin solid and dashed lines).

These oscillations are always damped until they disappear, with a damping rate
depending on the cavity detuning.
Precisely the oscillations are more persistent
the closer is the detuning to the bistable region.
This fact points to the competitive origin
of the oscillations. In Fig.\ref{monomode} (c) we show an example of the transient
dynamics when the cavity detuning is slightly moved from resonance.
It is worth to note that
the system is so sensible to noise that the duration of the oscillations suffers
slight variations, without apparent changes in the conditions.

Successively, we studied the response of the system
to small linear anisotropies driven by the tilt angle $\theta$ of the intracavity window W around
the vertical axis.

Several parameters of the system are simultaneously affected by this action.
As a consequence of the common increasing of the optical path for both polarization eigendirections,
the line center  moves with respect to the isotropic
condition $\theta=0$. In order to keep the bistable condition,
the cavity length is varied to recover the resonance.

In Fig.\ref{detuning_scan}((b) and (c)) we show the
intensity of both polarization components for different $\theta$ values,
showing that the gain profile is not significantly distorted with respect to the isotropic condition $\theta=0$
(Fig.\ref{detuning_scan}(a)). The detuning anisotropies are measured to be smaller than 5$\%
$ of the total change.

The other effect of the tilt angle $\theta$ is to increase the cavity losses.
In Fig.\ref{angle} (b) we report the dependence of the laser intensity
on the tilt angle $\theta$ which is in agreement with the expected theoretical behavior
\cite{Born}.
Moreover, since the gain profiles for the
two polarization directions do not change due to the tilt angle $\theta$ (Fig. \ref{detuning_scan}),
we can assume that the induced loss anisotropies remain a small effect ( $\simeq 1\%
$)
compared to the total loss change. The effect of the losses anisotropy can be observed in the residual modulation
in the total intensity (Figs.\ref{monomode} (a),(c)).

The dynamical effect observed when $\theta$ is increased
is the growing of the frequency of the polarization
oscillations. Measurements made near the resonance condition show that the frequency
rises
as the tilt angle increased(Fig.\ref{angle} (a)).
It can be seen that a tilt angle of 12$^o$
is enough to rise the frequency from 100 KHz to 400 KHz.

\section{Model and numerical results}
\label{s:model}

Our theoretical approach is based on the theory of the isotropic laser developed in Ref.
\cite{Puccioni87} where the optical coherences between
upper levels are considered with the. This theory was developed for the simplest
case (J=1 $\rightarrow$ J=0), while the transition involved in our system is 
much more complicated (J=19 $\rightarrow$ J=20). However, this theory
has been showed to predict also the behavior of lasers with a different
level structure, as shown in Ref. \cite{Puccioni89}. Only first
order coherences ($\Delta m=\pm 1$) will be considered. Therefore, independently of
the number of sublevels, there are only two kind
of possible transitions, which generate a split of the population in two ensembles,
in such a way that an anisotropy is induced in
the active medium \cite{Stephan98}. Furthermore we introduce 
an extrinsic linear anisotropy as done in Ref.\cite{Matlin95}.

The field will be decomposed in a circularly polarized base. Just losses and linear detuning
anisotropies along the principal axes
of the system will be included,  but not circular asymmetries since our system does not
show signs of dicroism. It reads as
\begin{eqnarray} \label{MB}
\dot{E_R} &=&\kappa (P_R-E_R)+i\delta E_R -(\alpha+i\beta)E_L, \nonumber \\
\dot{E_L} &=&\kappa (P_L-E_L)+i\delta E_L -(\alpha+i\beta)E_R, \nonumber \\
\dot{P_R} &=& -\gamma _{\perp }[P_R- D_R E_R - E_L C], \nonumber \\
\dot{P_L} &=& -\gamma _{\perp }[P_L- D_L E_L - E_R C^*],\\
\dot{C}   &=& -\gamma _c C - \frac{\gamma _ \parallel}{4}  (E_L ^* P_R + E_R P_L^*), \nonumber \\
\dot{D_R} &=&-\gamma _{\parallel }[D_R -r + \frac{1}{2}(E_R P_R^{*}+E_R^{*} P_R) + \frac{1}{4}( E_L P_L^{*}+E_L^{*} P_L)],\nonumber  \\
\dot{D_L} &=&-\gamma _{\parallel }[D_L -r + \frac{1}{2}( E_L P_L^{*}+E_L^{*} P_L)+\frac{1}{4}(E_R P_R^{*}+E_R^{*} P_R)] \nonumber
\end{eqnarray}
where $E_R(\vec{r},t),E_L(\vec{r},t)$ are the slowly varying electric fields
in the circular basis. $P_R(\vec{r},t),P_R(\vec{r},t)$ stand for
the matter polarization fields, $D_R(\vec{r},t)$ and $D_L(\vec{r},t)$  are the
respective population
inversions.  The  field $C(\vec{r},t)$ represents the coherence between the upper
sublevels. 
We recall that, in a density matrix treatment, the polarization correspond
to off diagonal matrix elements between upper and lower level of the radiative transition, whereas C is 
proportional to the of diagonal matrix elements coupling different angular momentum states of the upper
level \cite{Puccioni89}.
The parameter $\delta$ represents the detuning between the cavity
and the atomic transition frequencies. 
The parameters  $\alpha=(\kappa_V-\kappa_H)/2$ and  $\beta=(\delta_H-\delta_V)/2$
represent respectively the linear anisotropies in the losses and detuning with respect to the
cavity H-V axes, where $\kappa_V$, $\kappa_H$, are the losses
in the horizontal and vertical axis, and $\delta_V$, $\delta_H$ are the
corresponding detunings.

\begin{figure}
\centerline{\scalebox{1.3}{\includegraphics{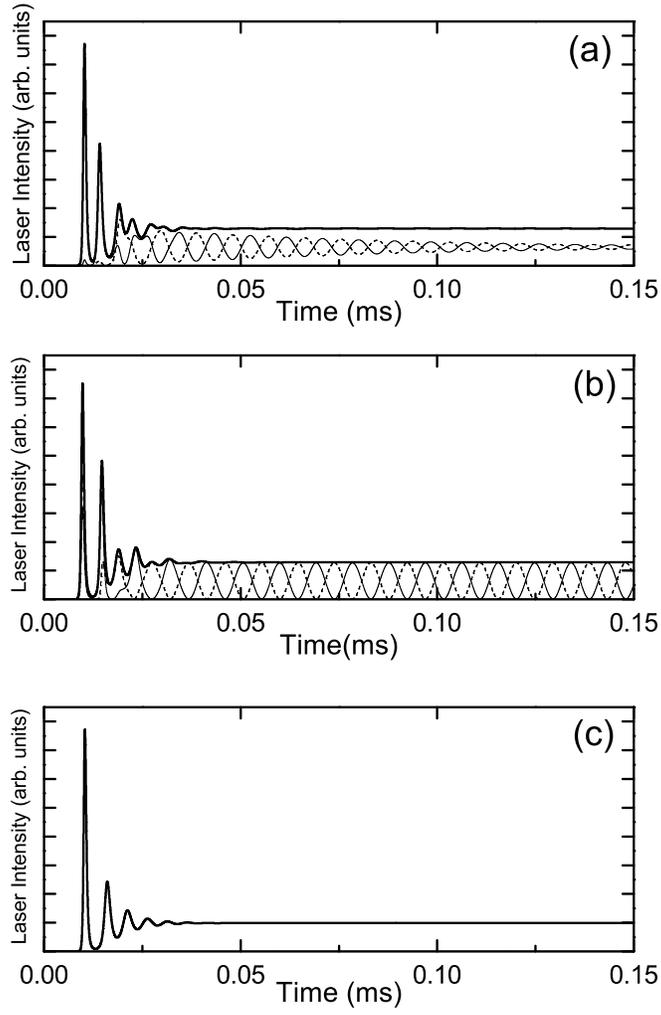}}}
\caption{Numerical intensity profiles for $\delta=0.05$, $\theta$=0 and :(a) $\alpha=\beta$=0.005,
 (b) $\alpha=\beta$=0.0 and  (c)  $\alpha=\beta$=0.005, C=0.}
\label{no_coh}
\end{figure}

In our low pressure CO$_2$ laser, the polarization decay is $%
\gamma _{\perp }=4.4\cdot 10^8$ s$^{-1}$ and the inversion decay rate as $%
\gamma _{\parallel }=1.95\cdot 10^5$s$^{-1}$ \cite{Leyva01}.
In view of numerical solutions, there is no reason to keep dynamical equations for
variables whose decay rates differ by several orders of magnitude. Depending on whether
$\gamma _c \simeq \gamma _\parallel$ or $\gamma _c >> \gamma _\parallel$, we can respectively
reduce the number of differential equations from 7 to 5 (for E$_R$, E$_L$,C, D$_R$ and D$_L$) or 4
(for E$_R$, E$_L$, D$_R$ and D$_L$). In the first case, P$_R$, P$_L$ are eliminated adiabatically and their values
correspond to the stationary solutions of the third and fourth of Eqs. \ref{MB}.
In the second case (fast C), also the C variable is eliminated adiabatically and it follows
the relatively slow variations of the field and the population inversion. Whether to choose between
five or four equations depends on the numerical value of $\gamma _c$.
The $\gamma_c$ parameter represents the coherence decay rate, whose value should be
chosen between $\gamma_\perp$ and $\gamma_\parallel$ \cite{Abraham96}. However, this
parameter can not be directly measured, and it will be used as
control parameter in order to fit the theory to the experimental results \cite{Matlin95}.
We find that the optimal value is $\gamma_c \simeq \gamma_\parallel$
in all cases, which is also consistent with the observation that just linearly polarized states are found in the experiment.
Indeed, a higher value of $\gamma_c$  would give rise to a periodic modulation of the
total intensity  \cite{Abraham96} which has never been observed in our  experiments.
Hence, the coherence C decays on the same time scale as the population inversion and we must keep the
corresponding time dependent equation.  Thus our experimental system is dynamically modeled as
a class B laser \cite{arecchi}, and Eqs. (1) reduce to the following system of two algebraic equations:
\begin{eqnarray}
P_R &=& D_R E_R + E_L C \nonumber \\
P_L &=& D_L E_L + E_R C^*
\end{eqnarray}
and five differential equations:
\begin{eqnarray} \label{adiabatic}
\dot{E_R}   &=& \kappa(D_R-1)E_R+i\delta E_R +\left(\kappa C -(\alpha+i\beta)\right)E_L , \nonumber\\
\dot{E_L}   &=& \kappa(D_L-1)E_R+i\delta E_L +\left(\kappa C^* -(\alpha+i\beta)\right)E_R , \nonumber\\
\dot{C}   &=& -\gamma _\parallel(C(1+\frac{1}{4}(|E_L|^2+|E_R|^2))-\frac{1}{4}E_R E_{L}^*(D_L+D_R)) , \\
\dot{D_R} &=&-\gamma _{\parallel }\left[D_R(1+|E_R|^2)
-r+\frac{1}{2}\left(D_L|E_L|^2+\frac{3}{2}(E_RE_{L}^{*}C^*+E_{R}^{*}E_LC)\right)\right] , \nonumber\\
\dot{D_L} &=&-\gamma _{\parallel }\left[D_L(1+|E_L|^2)
-r+\frac{1}{2}\left(D_R|E_R|^2+\frac{3}{2}(E_RE_{L}^{*}C^*+E_{R}^{*}E_LC)\right)\right] . \nonumber
\end{eqnarray}
Eqs. (\ref{adiabatic}) show explicitly that the cross field coupling 
is due partly to a molecular coupling $J=1 \rightarrow J=0$ (intrinsic) and partly to
the intracavity window W (extrinsic).
Fig. \ref{no_coh} provides evidence that the main responsible for the alternation 
is the intrinsic anisotropy rather than the extrinsic one.
Figure \ref{no_coh} (a) refers to the case where both
coherences and linear anisotropies are acting. In Fig. \ref{no_coh} (b), 
the extrinsic anisotropy has been removed.
now the oscillation competition is not damped, but the alternation of the two polarization is maintained.
Finally, in Fig. \ref{no_coh} (c) just the 
extrinsic anisotropies are acting; the competition does not appear any more.


From the experimental observations we know that the most important effect of the window tilting
is the increase of the total loss, while the losses and detuning
anisotropies represent a minor effect. In addition, as the excitation current does
not change during the experiment, the increasing of the
losses affects the gain to loss ratio.
As a consequence, the total losses $\kappa$ and  pump parameter $r$ are functions of
$\theta$ as follows  \cite{Born}:
\begin{eqnarray}
r(\theta)=r_o(1-a_r\sin(\theta)^4) \\
\kappa(\theta)=\kappa_o(1+a_k\sin(\theta)^4)
\end{eqnarray}
where $r_o=2$ and  $\kappa_o=-\frac c{4L}log(R)=4.1\cdot10^6$ s$^{-1}$ are the pump strength and total losses
when the tilt angle $\theta=0$. Here $c$ is the speed of light, $L=0.75$ m the cavity
length and $R=\sqrt{R_1R_2}=0.956$ the mirrors reflectivity.
the parameters $a_r=150$ and $a_\kappa=25000$ are related to the
reflectivity of the window W, and are chosen to fit the experimental results.

In Fig. \ref{monomode} (b) and (d) the numerically generated intensity profiles are compared
with their experimental counterparts reported in Fig \ref{monomode} (a) and (c), respectively.
It can be observed that the intensity profiles show antiphase oscillations for both
polarization components, while the total intensity remains unmodulated.

When $\delta$=0, the polarization oscillations remain undamped for any degree of anisotropy.
For $\delta \ne 0$, the oscillations are still undamped only in perfect cavity symmetry conditions  $\alpha=\beta=0$.
In Fig.\ref{monomode} (d) it can be seen that for $\delta=0.05$, a detuning or losses anisotropy of
0.5 $ \%
$ is sufficient  to damp the oscillation in a few hundred microseconds as observed in the experiment.
In the experimental system unavoidable residual anisotropies break the cylindrical symmetry, and therefore
 the polarization oscillations are always damped.

\begin{figure}
\centerline{\scalebox{1.0}{\includegraphics{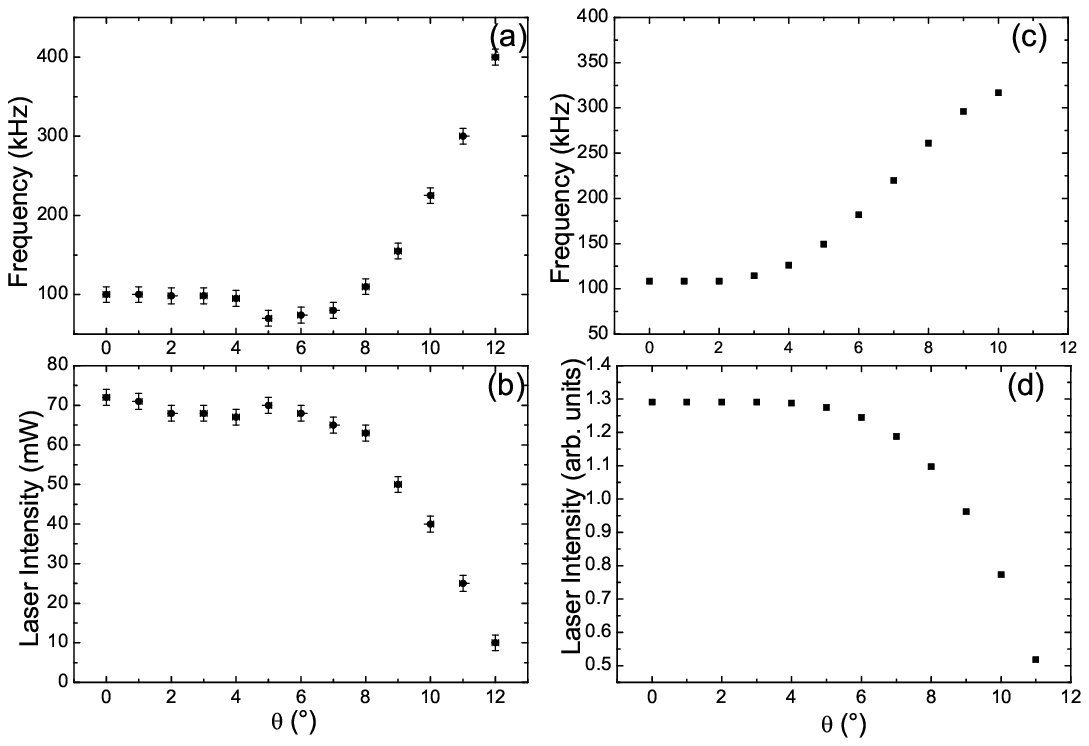}}}
\caption{Experimental angular dependences of: (a) polarization oscillation frequency
and (b) total intensity. Numerical dependences ($\delta=0$, $\alpha$=0.01 $\beta$=0.01):
(c) polarization oscillation frequency and (d) total intensity.}
\label{angle}
\end{figure}

Choosing the resonant condition $\delta=0$, we observe that the frequency of the oscillations presents the same
dependence on the total losses and pump strength found in the experiment.
In Fig. \ref{angle} we compare  the experimental and numerical angular dependences
of the polarization oscillation frequency (Fig. \ref{angle} (a),(c)) and the total intensity
 (Fig. \ref{angle} (b),(d)), showing good agreements.

\section{Conclusions}
\label{s:conclusion}

In this work we have reported an experimental and theoretical study of
the transient polarization dynamics of a quasi-isotropic CO$_2$ laser.
A competition between two modes with orthogonal polarization directions is observed. 
The competition manifests itself as an oscillation on the polarized laser field,
whose features depend on the cavity parameters.

To explain the observed transient polarization alternation, both intrinsic and extrinsic anisotropies must be 
considered; the intrinsic ones provide the alternation and the extrinsic ones provide damping and residual 
ripple of the total intensity. The numerical results fit well with the experimental observations, pointing out the
importance of matter variables in the dynamics of a quasi-isotropic laser.

\bigskip

The authors are grateful to F.T. Arecchi for  fruitful discussions.

\pagebreak

\pagebreak

\pagebreak

\pagebreak

\pagebreak

\pagebreak

\begin{thebibliography}{99}

 \bibitem{Lamb67} M. Sargent and W. E. Lamb. Phys. Rev {\bf 164},  436 (1967).

\bibitem{Bakaev87} D. S. Bakaev, V. M. Ermachenko, V. Yu Kurochin, V. N. Petrovshil, E. D.
Protsenko, A. N. Rurukin and Shananin. Sov. J. Quantum. Electron {\bf 18}, 1 (1988).

\bibitem{Puccioni87} G. C. Puccioni, M. V. Trantnik, J.E. Sipe and G. L. Oppo  Opt. Lett. {\bf 12},(1987).

\bibitem{Abraham95}  N. B. Abraham, E. Arimondo, M. San Miguel. Optics Comm. {\bf 117}, 344 (1995).

\bibitem{Abraham96}  N. B. Abraham, M. D. Matlin and R. S. Gioggia. Phys. Rev. A {\bf 53}, 3514 (1996).

\bibitem{Stephan98} For a review see: G. M. Stephan and A. D. May. Quantum Semiclass. Opt. {\bf 10}, 19 (1998).

\bibitem{Born} {\it Principles of Optics}. M. Born and E. Wolf. 6$^{th}$ edition. Cambridge University Press (1980).

\bibitem{Matlin95}  M. D. Matlin, R. S. Gioggia, N. B. Abraham, P. Glorieux and T. Crawford  Opt. Comm. {\bf 120}, 204 (1995).

\bibitem{Redondo97} J. Redondo, G. J. de Varc\'arcel and E. Rold\'an. Phys. Rev. A {\bf 56}, 648 (1997).

\bibitem{Vilaseca} A. Kulminskii, R. Vilaseca, R. Corbalan and N. B. Abraham, Phys. Rev. A {\bf 62}, 648 (2000).

\bibitem{Puccioni89} G. C. Puccioni, G. L. Lippi and N. B. Abraham.  Optics Comm. {\bf 72}, 361 (1989).

\bibitem{arecchi}  F.T. Arecchi in {\it Instabilities and chaos in quantum optics},
(Eds. F.T. Arecchi and R.G. Harrison), 
Springer Series Synergetics, Vol. 34 pp. 9-48 (1987).

\bibitem{Taggiasco97}  C. Taggiasco, R. Meucci, M. Ciofini and N. B. Abraham. Optics Comm. {\bf 133}, 507 (1997).

\bibitem{Labate98}  A. Labate, R. Meucci, M. Ciofini,  N. B. Abraham and C. Taggiasco. Quantum Semiclass. Opt. {\bf 10}, 115 (1998).

\bibitem{Leyva01}  I. Leyva, E. Allaria and R. Meucci. Optics Lett. {\bf 26}, 605 (2001).

\bibitem{Eijkele95} M. A. van Eijkelenborg, C. A. Schrama and J. P. Woerdman. Optics Comm. {\bf 119}, 97 (1995).

\end{thebibliography}
\end{document}